# Writing Across the World's Languages: Deep Internationalization for Gboard, the Google Keyboard

Daan van Esch,* Elnaz Sarbar, Tamar Lucassen,
Jeremy O'Brien, Theresa Breiner, Manasa Prasad,
Evan Crew, Chieu Nguyen, Françoise Beaufays
Google     Mountain View, CA, USA

November 2019

**Abstract**

This technical report describes our deep internationalization program for Gboard, the Google Keyboard. Today, Gboard supports 900+ language varieties across 70+ writing systems, and this report describes how and why we have been adding support for hundreds of language varieties from around the globe. Many languages of the world are increasingly used in writing on an everyday basis, and we describe the trends we see. We cover technological and logistical challenges in scaling up a language technology product like Gboard to hundreds of language varieties, and describe how we built systems and processes to operate at scale. Finally, we summarize the key take-aways from user studies we ran with speakers of hundreds of languages from around the world.

## 1 Introduction

Our world has a tremendous wealth of linguistic diversity, with thousands of languages spoken around the globe every day. Historically, most of the world's languages have by and large been used only in spoken face-to-face conversations, with very little writing taking place in the majority of languages. But as more of the world comes online, many language varieties that have mostly been limited to spoken usage in the past are now being used increasingly in writing to communicate using online message boards, chat apps, and social media, as described by e.g. Kral (2010), Osborn (2010), Kral (2012), Jones and Uribe-Jongbloed (2012), Pischloeger (2014), Nguyen et al. (2015), Keegan et al.

---

*Correspondence about this technical report can be sent to dvanesch@google.com. For product feedback or feature requests, please visit the Play Store page for Gboard at https://goo.gle/gboard.

(2015), Cru (2015), Lillehaugen (2016), Lackaff and Moner (2016), Jongbloed-Faber et al. (2017), Jany (2018), Soria et al. (2018), McMonagle et al. (2019), Eberhard and Mangulamas (2019), and McMonagle (2019).

In general, the trend seems to be that people want to communicate in informal environments like chat apps and social media in the same kind of language they would normally speak in face-to-face conversations. And since chat apps and social media are usually (but not always) used to communicate through text, many more of the world's languages are now being written regularly by their users; perhaps only in informal contexts for now, but the trend is unmistakably moving in the direction of more languages being written on an increasingly regular basis. Regular use of smartphone applications for informal communications appears to be amplifying the grassroots-literacy trends observed in works such as Blommaert (2008).

In this technical report, we describe how we have been bringing support for hundreds of such language varieties to Gboard, Google's smartphone keyboard for the Android operating system[1], in order to help smartphone users around the world communicate and share knowledge in their preferred language(s). Gboard supports more than 900 language varieties today. It is installed out-of-the-box on many Android smartphones, and for most other Android smartphones, the application can be downloaded from the Google Play Store. Overall, it has more than 1 billion installs worldwide. As this report will show, the hugely diverse pool of people using smartphone apps like Gboard means that language technology now needs to support many more language varieties than has historically been the case.

The focus of this report is not on the technical implementation of our keyboard application, which is described in Ouyang et al. (2017) and other technical papers. Rather, this report focuses on the implications of the complex interactions between today's global linguistic usage trends and language technology products such as keyboards. In short, the confluence of widely accessible technology and informal written communication platforms has led to demand for many more language varieties to be supported in applications such as smartphone keyboards. This report describes in some detail technical and non-technical challenges that language technology product development teams face when scaling up to hundreds of language varieties, and the solutions we invented along the way.[2]

First, we'll describe in some more detail the sociolinguistic background against which many of the world's language varieties are now increasingly written, as well as some of the technological challenges encountered by language communities when doing so. Users have generally responded to these challenges by inventing various work-arounds, which we'll catalog briefly. Then, we'll de-scribe how we have been going about building support for many more language varieties in our smartphone keyboard application. Finally, we will present an analysis of some of the usage trends we are seeing, and provide an overview of future challenges to be solved.

---

[1] Gboard is also available on the iOS operating system, supporting a subset of the language varieties available on Android.

[2] Throughout this report, ISO 639 language codes are formatted like `en` or `nan`, where two-letter language codes are used whenever available. Four-letter codes like `Latn` are ISO 15924 script codes. Character names from the Unicode standard are formatted like LATIN SMALL LETTER A.

## 2 Background

### 2.1 Languages, Writing, and Technology

Historically, writing was commonplace only for language varieties that were used in formal written publications, like books, newspapers, and religious materials. Today, however, many language communities have access to chat apps and social media, which are generally informal environments. Most communication in chat apps and on social media happens in writing, but language usage is more similar to the patterns that would historically have been restricted to spoken usage (McCulloch, 2019). Along these lines, to communicate in natural ways within this informal yet mostly written context, many speakers across the world have picked up writing in their own language varieties, even if these language varieties were rarely written historically.

In general, however, support for these language varieties without a long-standing widespread written tradition remains rare within language technology products, despite early efforts to address this problem, as described in e.g. Paterson (2015). This means there are significant opportunities for language technology products to make a positive difference for their users by adding in support for many more language varieties around the world. To cite just one example, the Digital Language Diversity Project, which studied a number of European regional languages, calls out "technological barriers, such as the unavailability of a specific keyboard or spell checkers that would ease the writing" as one of the main problems facing the languages communities they studied in the use of their languages online (Soria et al., 2018).

Typically, chat apps and social media are accessed using smartphones. Text input on these devices is generally facilitated by a virtual keyboard application, displaying a keyboard layout (such as QWERTY or AZERTY) on-screen and using the touchscreen capability to detect taps or gestures. Because these screens are small, machine-learning language technologies like predictive text and autocorrection can help make input faster and more accurate (Fowler et al., 2015; Ouyang et al., 2017). Until about 2016, these technologies have only been available in about 100 language varieties.

In the last few years, however, it has become very clear that, with the rise in informal written communication globally, along with the increasing ubiquity of smartphones (ITU/UNESCO Broadband Commission, 2017), language technology needs to scale beyond the traditional set of about 100 language varieties in order to support communities across the world. Wikipedia, for example, is already available in about 300 language varieties (Wikipedia, 2019). Beyond Wikipedia, Scannell (2007) and our previous research (Prasad et al., 2018) found textual data in more than 2,000 language varieties online.

We propose calling efforts to bring technology to many more languages at scale 'deep internationalization', after the industry-standard term 'internationalization'[3], which is typically used to mean extending support to languages and communities beyond American English, the language and locale that most software products are designed to support first. However, the term 'internationalization' has not typically been thought of as extending to hundreds or even thousands of language varieties, so we added the adjective 'deep' to distinguish our effort from more limited internationalization efforts.

## 2.2 Linguistic Diversity Online

Over the last few years, we have been working on a deep internationalization project to add support for many more languages to Gboard, our smartphone keyboard application. Before we decided to start working on deep internationalization for Gboard, we conducted a number of literature surveys, data analyses, and observational studies to understand user expectations around the use of language in the smartphone age.

Earlier in this report, we already referenced a number of papers analyzing the general usage trends for these language varieties, e.g. in analyses of Twitter posts. It stands to reason to assume that use of these language varieties would be even more common in less public spaces than Twitter, such as semi-private community groups on social media platforms, and private chat apps with small group conversations. Some language-specific evidence for this assumption is found in Jongbloed-Faber et al. (2016), but it is hard to prove this assumption conclusively on a global scale.

However, some additional signals for the linguistic diversity of the digital world can be seen in projects where communities created their own user interface translations for popular social media sites (Scannell, 2012), following a trend of creating localized input methods and user interfaces for open-source operating systems such as Linux (Reina et al., 2013).

Another data point is that even back in 2016, when we started this project, there were already hundreds of languages with an active Wikipedia edition. The text in Wikipedia articles is generally well-edited, in addition to being labeled with a standard ISO 639 language variety tag (with some unfortunate exceptions, such as `als` being used for Alemannic instead of Tosk Albanian), so it is often quite suitable for deriving candidate wordlists for spell-checking purposes. It is worth pointing out that having a dedicated Wikipedia is not, in and of itself, conclusive evidence of wide-spread use by language communities, since the pool of editors for some of these editions is relatively small, and the content of some of these editions appears to have been mostly generated automatically by bots, but all the same, it is encouraging to see so many languages represented online in Wikipedia.

---

[3]Also known as 'i18n', pronounced 'i-eighteen-n', with '18' representing the number of letters in between the initial 'i' and the final 'n' in 'internationalization'.

## 2.3 Making Do Without Tailored Language Technology

### 2.3.1 Character Sets, Writing Systems, and Keyboard Applications

In our user studies, we observed a number of creative work-arounds for problems that language technology was putting in the way. For example, in Kanuri (`kr`), spoken in the region around Lake Chad in Western Africa, some speakers use the digit '3' to stand in for the letter 'ə' (LATIN SMALL LETTER TURNED E), due to 'ə' being unavailable on standard English keyboard layouts, and the unavailability of a language-specific layout with 'ə' included. In other languages, we observed users omitting the diacritics the regular orthography would use; this was not always because users preferred to skip these diacritics, but frequently also because they were simply unavailable on the virtual keyboards these users were using, as they were designed for English.

Of course, many language varieties are not written in the Latin/Roman alphabet that is used for English. High-quality input methods have long existed for some other writing systems, such as Cyrillic (`Cyrl`), Chinese (`Hans`, `Hant`), Japanese (`Jpan`), and Korean Hangul (`Hang`). But even though smartphone keyboards have long supported the Cyrillic alphabet in general, Cyrillic-script layouts were typically restricted to the character sets used in large languages like Russian. This meant that, even though the Cyrillic script was generally thought of as well-supported in technology, for many years, speakers of e.g. the Yakut language in Siberia (`sah`) still faced issues with characters they needed being unavailable in their keyboard layout, specifically 'ҕ' (CYRILLIC SMALL LETTER GHE WITH MIDDLE HOOK), 'ҥ' (CYRILLIC SMALL LIGATURE EN GHE), 'ө' (CYRILLIC SMALL LETTER BARRED O), 'һ' (CYRILLIC SMALL LETTER SHHA), and 'ү' (CYRILLIC SMALL LETTER STRAIGHT U), and their capital-letter equivalents. Examples like Yakut and Kanuri make it clear that keyboard layouts need to be tailored to each language variety, even for languages written in scripts that already have existing keyboard layouts.

While input methods have been reasonably widely available for Latin, Cyrillic, Chinese, Japanese, and Korean, input methods for many other writing systems have been harder to find. This has resulted in language communities creating 'online' orthographies for languages written in these writing systems, typically using the Latin alphabet instead of the standard writing system.

One well-known example is Arabizi, where Arabic (`ar`) is written using Latin letters and digits, e.g. 'keif 7alak?' rather than the standard Arabic script, i.e. 'كيف حالك؟' in this case. More details on Arabizi can be found in Bjørnsson (2010) (which also explains how the emergence of Arabizi is also related to the wide gap between standard formal Arabic writing and conversational everyday Arabic speech).

Similarly, many languages of the Indian subcontinent, such as Hindi (`hi`), now also have a widely used 'online' Latin-script orthography, which has emerged alongside the native writing systems used for the standard orthography, such as Devanagari (`Deva`) in Hindi (Wolf-Sonkin et al., 2019).

In some of these situations, our user studies suggest these 'online' orthogra-phies carry additional symbolic connotations, such as an association of the Latin alphabet with English leading to these new orthographies being seen as trendier than the standard writing system, as in Hindi. In these cases, users may well prefer to keep using this 'online' orthography, and it may make sense to support this newly emerging orthography directly in a keyboard application as well, as Gboard does for a number of Indian languages. Even so, we do believe that language communities should have the option to use a high-quality input method tailored exactly to their language variety in its standard writing system.

### 2.3.2 Linguistic Typology and Language Technology

During our research, we observed a number of interesting characteristics of 'online' orthographies, such as the use of the digit '2' in languages like Indonesian (`id`) to mark a common morphological construction called reduplication, meaning the (partial) repetition of a word to achieve a different meaning. This phenomenon is rare in English, but can be observed in the sentence pair 'I like you' vs. 'I like-like you', where the second version, with the reduplicated verb, indicates romantic interest unambiguously.

Reduplication is very common in Indonesian and its linguistic relatives, and even though many smartphone keyboard applications already offer support for Indonesian, next-word prediction algorithms do not make special allowances for reduplication, meaning users would have to re-enter the same word all over again. Rather than re-typing the previous word from scratch in many cases, it seems that many users prefer to just type '2', as in e.g. 'makan2' instead of 'makan-makan'. This can be seen as another creative solution to language technology that isn't quite tailored to the specifics of the language at hand. To the best of our knowledge, a systematic way to handle reduplication and similar morphological phenomena in n-gram language models is yet to be developed, and we believe it would be a fruitful area for future work.

More generally, even typological phenomena such as reduplication, which are well-known to linguists, can be challenging to handle using many standard approaches, such as n-gram models. Beyond reduplication, linguists would be quick to point out that many more phenomena exist that equally pose challenges for language technology, such as agglutination and polysynthesis (Littell et al., 2018; Mager et al., 2018). These phenomena appear in many of the world's languages, and it seems like an important area for future work to support them better in language technology, although it is also important to mention that these terms may be defined too broadly to be practically applicable (Haspelmath, 2018). We think it would also be fruitful to investigate further what kind of a user experience may be ideal for users whose languages feature such typological phenomena: for example, perhaps instead of having a next-word prediction feature, there should be a next-morpheme prediction feature.

### 2.3.3 Linguistic Variation and Language Technology

Even for language varieties whose writing systems and character sets have long been readily supported by smartphone keyboards, users may encounter challenges due to the lack of technological support for their specific variety. For example, speakers of Frisian (`fy`) and Limburgish (`li`), spoken in the Netherlands by 500,000 and 1,000,000 speakers respectively, would generally be able to access all relevant characters and diacritics by just configuring their smartphone to use a standard Dutch (`nl`) keyboard layout. However, spell-check, auto-correction and next-word prediction would then operate in Dutch, not Frisian or Limburgish. This means that most Frisian or Limburgish words would be auto-corrected wrongly by the underlying Dutch language models, or underlined as spelling mistakes due to their absence in the Dutch spell-checking wordlists.

Our research showed that such situations are commonplace the world over—speakers of a given language variety can frequently access the writing system and the character set they need via a similar language variety, but this means that other features such as auto-correction and spell-checking are mismatched with the target language variety.

To give just one other example, many users in the Arabic-speaking world, even when using a keyboard layout designed for Modern Standard Arabic, told us that most chat and social media messages are handled incorrectly by the auto-correction and spell-checking features built into their keyboard, because they typically write such messages in colloquial local varieties of Arabic, not Modern Standard Arabic.

We found that speakers, in general, adopt one of two approaches when faced with such mismatches. Most commonly, users appear to simply address such situations by continuing to type in their own language variety while teaching the on-device personal dictionary in their keyboard software all the words they need: they do this by reverting any incorrect auto-corrections manually, and adding words to the spell-check dictionary as they go.

Another approach is that users simply go into the settings and turn off all smart features, sticking with just the target layout, taking away the inconvenience of having to revert lots of inappropriate auto-corrections manually. In these situations, these users can input faster than when auto-correct is operating in the wrong language, but they will not benefit from the additional typing speed improvements that they would see if they were to use auto-correct and spell-checking in their preferred language variety.

### 2.3.4 Avoiding Language Technology Altogether

To wrap up, we will point out a few other approaches that we observed or heard about. One relatively common approach, especially in some languages of East Asia, such as Chinese varieties, appears to be sending voice recordings rather than typing out messages in writing. Such voice recordings are typically sent through private chat apps, not shared on public social media, and it is unclear how frequently users elsewhere use this approach. However, based on anecdotal evidence from East Asia and India, we believe it may be very common, particularly in low-literacy language communities.

Finally, one creative approach, which we believe to be rare, but which does illustrate the great challenges that the absence of language technology can pose for communities: some speakers of West African languages told us that, because they could not type in the writing system they generally used for their language, they were simply hand-writing notes, then taking photos of these notes, and sending these photos to each other via chat apps - thus avoiding the need for digital language technology altogether, but at a high cost in terms of convenience and data transferred, with digital photos being much larger to transmit than Unicode-encoded text.

## 3 Creating a Language Roadmap

Once we had concluded that we wanted to add support for many more language varieties to Gboard, the first question we asked ourselves was which ones we should add, and in what order. Creating such a language roadmap turns out to be a relatively challenging task even in itself, given the sheer linguistic diversity of our world, and the large number of factors involved.

Even just arriving at a precise count of our world's inventory of languages is generally agreed to be an impossible task, with unclear demarcations between similar language varieties, and lots of linguistic variation even within languages generally seen as one unified formalized language: for example, even English has many regional varieties, and there are also varieties such as African-American English. Even if we simply assume that there are thousands of language varieties in our world today, without attaching a precise number, it's clear that determining which varieties to add, and their relative prioritization, is a complex task.

### 3.1 Which Languages to Include

As an ideal, one goal might be for all language technologies to support all languages. However, we should recognize that in this particular context of building a smartphone keyboard application, it does not necessarily make sense to support all 6,000-7,000 languages at the present moment.

Even considering the trend towards writing many more languages online which we described above, perhaps half of the world's language communities have yet to develop (and some may never develop) even informal, 'online' orthographies. These languages may be used by groups where:

- literacy rates remain low

- technology is not yet accessible or available

- there is simply no demand (yet) for writing in their language, and perhaps no orthography exists at all yet

In these situations, speakers would be unlikely to benefit from the creation of a keyboard application for the language variety at this point in time: they would not necessarily use it, and it will probably even be unclear what characters should be included in the layout. In many situations, developing other language technologies may make more sense: for example, automatic speech recognition technology may be a better fit than a keyboard application. And of course, some of the world's languages are sign languages, where different kinds of technology may be needed.

As a general philosophy, we think it is important to focus on what would be most helpful for the users of each of the world's language varieties, and to proceed from there, while also understanding that the needs of these users are not homogeneous. We also think it is important to re-evaluate the situation regularly, since the situation on the ground may shift over time.

It is also worth pointing out that language communities can, and increasingly do, develop their own third-party Android keyboard applications or other types of language technology, frequently working with fieldwork linguists or other academics collaborating closely with these communities, using the kinds of approaches described in Paterson (2015). Thus, even if a language is not yet included in Gboard, it is still possible for an Android keyboard to be created and distributed for use among the community.

## 3.2 Prioritizing the Roadmap

Even so, the set of language varieties that needs to be considered for inclusion would still be quite large, so we had to create a prioritization system, allowing us to create a roadmap for language support. To be clear, of course, we believe all language varieties are equally valuable, but it is simply not feasible to work on hundreds of varieties simultaneously, so a rough prioritization is needed, maximizing the impact that can be achieved within the resources available. There are a number of factors that can be used in deciding the relative prioritization of each language variety, including:

- Can we find evidence of this language already being written online, even if just in informal contexts? Approximately how common is this, as far as we can determine? Is the language present in major multilingual sites, such as Wikipedia?

- Are there books, newspapers, magazines, or other formal written publications available?

- What's the approximate number of speakers?

- What do smartphone usage trends look like today, and what are the forward-looking projections?

- Are all the other i18n building blocks (Unicode encoding, fonts, etc.) in place for this language variety?

- Have we received feature requests for this variety to be added through various product feedback channels?

- What alternatives do speakers of this language currently have, e.g. is a layout for another language available that they may be able to use to input all relevant characters (even if they would have to disable the keyboard's auto-correct functionality)?

- Is it an official, administrative language of any country/territory?

When building roadmaps, some of these factors can be fetched from industry-wide resources such as the Unicode Common Locale Data Repository (CLDR), while others require in-house research. Collating this data makes it possible to bucket the language varieties based on these various factors. Then, we can work on a set of varieties with approximately equivalent priorities in parallel.

In our work, the presence of other i18n building blocks was only a factor to a limited extent, as virtually all relevant scripts were already included in the Unicode standard, and had fonts available through Google's Noto efforts (`https://google.com/get/noto`). The main technical consideration was whether most Android smartphones included an appropriate font for the writing system at hand, such as Google's Noto font for the script. However, the other questions frequently required significant research.

### 3.2.1 Uncertainty about Statistics

For example, beyond the academic studies in a handful of languages cited above, there is virtually no public information on how frequently most of the world's language varieties are used in e.g. private chat apps or in semi-private social media communities. And even determining seemingly simple statistics such as approximate speaker numbers is a tough challenge for many language varieties, with various sources, such as the English-language Wikipedia, Eberhard et al. (2019), and official government censuses all reporting wildly inconsistent numbers.

Certainly, in many cases, our roadmap planning involves making rough estimates, as much of the data that we would need to determine these numbers with any degree of certainty is unavailable. Given all this uncertainty, roadmaps are by necessity imperfect, but the goal is to sort the languages of the world in roughly the right order to maximize the impact and helpfulness of our work, given the resources available, and we find that this method generally achieves that goal.

## 3.3 Writing Systems and the Roadmap

In creating the roadmap, one complication we had not anticipated was determining the appropriate writing system(s) to support in the keyboard for each language variety.

### 3.3.1 Languages of India

For example, multiple writing systems are commonly used for many languages of India, such as Santali (`sat`), one of the 22 languages listed in the 8th Schedule to the Indian Constitution. Santali is written in a number of writing systems, including Devanagari (`Deva`), Bengali/Assamese (`Beng`), Ol Chiki (`Olck`), and Latin (`Latn`). In this case, it required a significant amount of research to determine which of these writing systems is more commonly used, and which of these writing systems Santali speakers would want to use. These two questions do not necessarily result in the same answer, given that the choice of writing system is frequently informed by the availability of technologies like keyboard applications.

Santali is by no means the only language of India for which multiple writing systems are in use. As described by Brandt (2014), in India, the choice of writing system for Indian languages is influenced by many extra-linguistic factors, and we find that generally, extensive research is necessary to determine which writing system(s) to use for most languages of India.

Beyond native writing systems and standard orthographies, as mentioned above, many Indian users express a strong preference for the use of the Latin alphabet for languages such as Hindi (typically written in Devanagari), even when presented with a fully functional keyboard for the standard writing system. As a result, we decided to offer support for Latin-script input for many Indian languages, with output either in the standard script through transliteration (Hellsten et al., 2017) or in the Latin alphabet (Wolf-Sonkin et al., 2019).

### 3.3.2 Indigenous scripts in South-East Asia

India is by no means the only country in which we face a complex landscape in terms of selecting which scripts to use for a given language variety. For example, in Indonesia, many languages have a historical tradition of being written in local scripts, such as Sundanese (`Sund`), Javanese (`Java`), Balinese (`Bali`), and Buginese (`Bugi`). These indigenous scripts have now largely been replaced by the use of the Latin alphabet in everyday usage, but they are still found in public signage, and form an important aspect of the cultural heritage (Kuipers, 2003). We generally choose to add support for the everyday writing system first, since that will be most helpful to people using our keyboard application, and potentially add support for lesser-used heritage scripts later. In Indonesia, for example, we added support for all four scripts mentioned here, but only after we had already added support for these languages in the Latin script.

In the Phillipines, the situation is somewhat similar: the everyday writing system is Latin, but heritage scripts such as Baybayin (`Tglg`) also remain in use. Again, we added support for the most commonly used writing system first, and later on, we added support for Baybayin, as well as three other indigenous scripts of the Phillipines (Tagbanwa `Tagb`, Hanunuo `Hano`, and Buhid `Buhd`).

### 3.3.3 Use of the Arabic script

During our research, we learned that in the Indonesian archipelago, historically, the Arabic script (`Arab`) was used for many languages spoken along trade routes, such as Acehnese (`ace`) (Daud, 1997). While the Arabic script has fallen out of use in most places in modern Indonesia, it continues to be used to some extent for Malay, although it may also be dwindling there (Yaacob et al., 2001). We have added support for some Arabic-script keyboards for languages of Indonesia and Malaysia, but again, we added support for the more commonly used Latin-script orthographies first.

Beyond Indonesia, many language varieties of Western Africa also continue to be written in the Arabic script in addition to the Latin script, due to historical influences in the region (Souag, 2011).

### 3.3.4 Multiple Orthographic Standards

Even for language varieties with just one writing system in common use, varying orthographic standards may still be used: in fact, this happens even in English, where the United States and the United Kingdom, for example, use slightly different orthographic standards. In many languages, the differences between varying orthographic standards are significantly larger, and the decision of which orthographic standard(s) to include can be challenging. Sometimes, it was clear that only one orthographic standard was still in everyday use, while in other cases, speakers of the target language are still debating which standard to adopt. These situations require a nuanced, tailored approach for each language variety. We found Oko (2018) a helpful case study to understand the complexities that can be at play; Jones and Mooney (2017) and Cahill and Rice (2014) also offer a wealth of knowledge.

### 3.3.5 Useful Resources

In the end, we primarily based the decision of which writing system(s) to include on on-the-ground usage factors, and on feature requests we received. Useful sources we consulted included Daniels and Bright (1996), the Unicode Standard, and related documentation (such as proposals to the Unicode Technical Committee, for which we found SIL International (2019) to be of immense value). We made extensive use of aggregators of linguistic research such as Hammarstrom et al. (2019). We also made use of Eberhard et al. (2019). Finally, simple overview charts with all relevant graphemes, as found in many language-specific Wikipedia articles, as well as on Omniglot (Ager, 2019), were tremendously useful to us.

## 3.4 Automatic Dashboards

Once we had decided which language varieties we would be supporting, in roughly what order, and in which writing systems, we needed a way to track our progress. Given the number of languages involved, manual tracking solutions such as spreadsheets would not be ideal: with dozens of sub-tasks per language, there would be many thousands of sub-tasks to track, and prior experience taught us that manually maintained spreadsheets would inevitably get out of date very quickly.

To give team members access to up-to-date information, we created a set of dashboards which automatically take in status information from our source control repository, such as whether a layout design has already been committed into the repository. We created dashboards for each of the major sub-tasks per language, so we can easily see which languages would need to be worked on next for each of these categories. For each sub-task to be executed for a given language variety, our dashboard automatically links in a pointer to detailed, step-by-step documentation on how to complete this sub-task. We also integrated this dashboard with our internal issue tracking system, so we can easily see if anyone is already working on an individual sub-task for a given language variety (tracked with an issue in our system), and if so, who; a critical feature in a globally distributed team.

This may appear as a nitty-gritty description of task management that would not typically warrant being described in detail, but it is interesting to consider that without the work that went into these automatic dashboards and our process documentation, the logistical challenges involved in adding support for hundreds of language varieties would have been daunting, presenting a nearly insurmountable obstacle to the large-scale deployment of language technology - and one that is in no way due to any complexity innate to the technology itself.

More generally, it strikes us that non-technical issues (such as scaleable task management systems, process documentation, unavailability in a standardized format of even the most basic information on orthographic systems for many of the world's languages, and limited availability of trustworthy standardized data on linguistic usage patterns across the world), in aggregate, are much bigger factors than many realize in contributing to the limited level of support in language technology for the world's languages today.

# 4  Designing Keyboard Layouts

In our experience, the most challenging aspect of our deep internationalization efforts for Gboard has been designing and implementing keyboard layouts across language varieties and writing systems. Virtual on-screen keyboard layouts have many benefits over physical keyboards, including that they:

- are easily adaptable for each language variety, with mere software configuration

- can contain more rows and columns than a physical layout

- allow designers to create dynamic layouts, where keys can change appearance after a key has been tapped—this helps account for complex writing systems with more graphemes than can comfortably fit on a single screen

- enable per-user layout personalization, e.g. by making it possible to enable or disable a permanent number row at the top of the layout

- offer the option to switch easily from one keyboard layout to another, e.g. for users who are familiar with multiple writing systems

All the same, it can be challenging to design usable layouts that can fit the necessary characters for each language into the limited amount of on-screen real estate that is available. All the characters that are required for a given language need to be placed in positions that make sense to the target user base, but at the same time, the overall space that the keyboard takes up should not cover too much of the device's screen. In many cases, even with dynamic layouts, it is practically impossible to put all required characters on the first page in the keyboard layout, and we had to decide which characters to make available as long-presses (i.e. keys which become accessible as their host key is held down for some longer period of time), and sometimes even which keys to put on a second or even third page.

While there are plenty of challenges when manually designing layouts across dozens of different scripts, we found that these factors made even seemingly simple Latin-script languages difficult to support at scale. We originally designed each layout manually, with linguists or native speakers starting out from a basic layout grid (such as QWERTY, QWERTZ, or AZERTY), then adding in long-press characters. To some extent, of course character frequency is the main consideration in deciding which long-press characters to include, and so given a text corpus, it seemed to us like this part of the layout design process could be automated. We have described our approach for doing so for Latin-script orthographies in Breiner et al. (2019).

Although we can employ automation to accelerate our layout design process for Latin-script orthographies, even layouts generated through this automatic system still require some human input to ensure they are laid out in the best-possible way, while languages with non-Latin script orthographies virtually always require extensive human input. The reasons human input is required for even Latin-script layout design are manifold, but include:

- The basic layout grid of choice for a given community is typically influenced by complex historical and cultural considerations. In many languages, familiarity with key placements of layouts for major lingua francas influences users' preferences for key placements in their native language: for example, if users have grown used to typing on a QWERTY layout when writing in English, or when writing their native language using an English keyboard layout, they tend to prefer to stick to the layout they know. It seems very challenging to design an automatic system that can accurately decide whether a given language community would prefer QWERTY, AZERTY, QWERTZ, or yet another basic layout grid.

- In some cases, different countries may have adopted different standard layouts for typing in their languages, even if these languages use the same writing system. For example, Canadian French uses a QWERTY layout by default, unlike French of France, which uses an AZERTY layout.

- Frequently, entirely new keys need to be added to the basic layout grid, as in the layout for German of Switzerland, which has keys for 'ä', 'ö', and 'ü' to the right of the standard QWERTZ grid. Without human linguistic input, it is not possible to determine whether such keys need to be included as stand-alone keys, or whether it would be sufficient to include them as long-press keys.

- In many cases, existing official layout standards exist, but these are not typically machine-readable. Even if such standards exist, there may be locally-developed keyboard apps that do not follow these standards, but that implemented a different keyboard layout instead that many users in the community have grown accustomed to.

- Even if the standard orthography for a given language does not include some letters, these letters may still be needed to write commonly used loanwords from neighboring or major world languages.

In languages using non-Latin scripts, the type of writing system significantly influences how layouts are designed. In some cases, the same principles outlined above apply equally, as in languages that use the Cyrillic script (`Cyrl`) or the Greek script (`Grek`). Similarly, for many abjads, such as the Arabic (`Arab`) script, a simple non-dynamic one-page layout is sufficient. However, for many abugidas, such as the Brahmic scripts of India (e.g. `Deva`), we use dynamic layouts extensively to facilitate typing all relevant characters on just one page. Similarly, in syllabaries such as Cherokee (`Cher`), we use dynamic layouts to make all necessary characters easily accessible.

In some cases, when developing dynamic layouts for abugidas and other complex scripts, very specific ordering is required for the rules that govern which keys are shown dynamically, in order to achieve proper rendering of the characters. Dynamic keys may remain blank until the relevant rules are activated, when the character combination is shown on the key. This allows for only licit characters to be combined dynamically and presented to the user. Without these combination rules, illegal or nonsensical character combinations would often appear on the keys, creating clutter and causing issues for the user when typing.

Another interesting factor is that some non-Latin script layouts are influenced by well-known Latin-script layouts such as QWERTY, leading to designs where similar graphemes from, say, the Cyrillic script are placed in the same positions on the layout as the equivalent graphemes would be in a Latin-script QWERTY layout.

All these layout design considerations mean that it typically requires a significant amount of human linguistic input to design a well-functioning layout. Overall, Gboard today includes 900+ layouts across over 70 scripts. Even if all the layouts in each script typically share some basic characteristics, each and every single one of these layouts has required a significant amount of linguistic work to tailor it to the individual language variety it will be used for.

# 5 Building Language Models

## 5.1 Gathering and Normalizing Data

Supporting features such as auto-correct, next-word prediction (predictive text) and spell-check requires the use of a machine-learning language model, such as n-gram language models, which can be used in a finite-state transduction decoder (Ouyang et al., 2017). These language models can be created based on a variety of textual sources, e.g. web crawls, external text corpora, or even wordlists (to create unigram language models). A detailed description of our standard approach to mining training data for language models across many languages can be found in Prasad et al. (2018). Since the data that we mine can be quite noisy, we apply our scalable automatic data normalization system across all languages and data sets, as described in Chua et al. (2018). Our model training algorithms are described in Allauzen et al. (2016).

### 5.1.1 Eliciting Additional Text Corpora

Where text corpora or wordlists are unavailable, or not sufficiently large to create a decent wordlist, we decided to work with language communities to create small text corpora. Our elicitation approach for such corpora is to provide prompts for native speakers to respond to in writing, such as "What was your favorite food growing up? Describe how you prepare it and why you like it." Our experiments showed that about 10,000 sentences is typically sufficient to train a small language model with a sufficient level of quality to handle basic auto-correction and spell-checking, so we created 5,000 unique prompts across different domains, gathering two responses for each of the prompts.

Creating corpora using such free-writing prompts generally works well, but one challenge that comes up from time to time is that for some language varieties, the native speakers we work with do not have access to any other keyboard, even on desktop or laptop computers, with the right writing system and character inventory. In those cases, a quick web-based virtual keyboard can be put together. In yet other cases, non-Unicode font encodings may be used in the data we receive from native speakers, necessitating the creation and/or use of a converter.

Another interesting challenge with these free-writing prompts was that the prompts we designed did not necessarily make sense around the world. We did our best to create a set of prompts that would apply across cultures and countries, but even though we had paid careful attention to this topic, we've still received notes from some native speakers pointing out that some questions do not make sense for them. For example, some native speakers pointed out that the question "What are the ten biggest cities in your country, and what are some of the best-known places in each of them?" did not make sense in their country, given that there weren't quite that many cities in their country. This once more emphasized the importance of having diverse perspectives throughout the development process.

## 5.2 Linguistic Variation and Language Models

While in general, creating a language model for given linguistic varieties using the approaches outlined above works quite well, in some situations, more complications arise due to the sheer amount of linguistic variation in the world. In the extreme, one could say everyone speaks every language variety that they speak in a slightly different way from all other speakers of that variety: people have their idiolects, meaning they have their own individual linguistic preferences and usage patterns, differing in terms of word choice, topics, and loanword sources, to name but a few sources of individual variation in language use.

Generally, though, such differences can be modelled with on-device personalization towards the user's individual language usage, with the core language being modelled by a generic model built for the language variety at hand, and on-device personalization taking care of the rest (Fowler et al., 2015).

However, some language varieties have a large degree of internal linguistic variation, e.g. Romansh (`rm`) of Switzerland, and Limburgish (`li`) of the Netherlands. In such language varieties, providing high-precision auto-corrections is challenging due to local orthographic and lexical variation, and such variation also complicates providing accurate next-word predictions.

Of course, one approach could be to simply handle each version of Romansh and Limburgish as an individual variety, building separate models for each and every one of them. This is, however, hardly practical, because enumerating and demarcating these varieties is quite challenging, and even if this could be done, handling each variety individually would also require a large set of language models to be built.

The approach we have adopted is to build one model covering all sub-varieties for each language variety, where we tune the auto-correction parameters to be significantly more lenient, and then to rely on on-device personalization to learn the user's individual preferences. Similar approaches can be adopted for many languages the world over, including colloquial Arabic varieties, which also offer a wide range of internal linguistic variation with unclear boundaries between varieties (Abdul-Mageed et al., 2018).

In fact, such personalization approaches may even make sense in languages that would not normally be considered to have a large degree of internal linguistic variation. For example, in English, there is a widely used standard form that is commonly used in writing (i.e. the form this report uses), but there is also a large degree of linguistic variation in English (Labov, 2012; Trudgill, 2016) that makes its way into informal writing, as shown by e.g. Eisenstein (2013) and Nguyen (2019). On-device personalization can help account for such differences in English, too. In future research, we hope to explore further how and when on-device personalization can help model fine-grained linguistic variation.

Of course, even for an individual language user, different degrees of formality and different linguistic registers may be used depending on the context (e.g. writing a business email vs. having an informal chat with a friend), so beyond personalization, investigating contextualization also appears to be a fruitful area for future work.

### 5.3 Multilingual Input

Many people around the world speak multiple languages. For example, in Europe, more than half of the population speaks more than one foreign language in addition to their native language (European Commission, Directorate-General for Communication, 2012). In most countries in Africa, multilingual speakers also form a majority of the population (Logan, 2018). Reliable statistics for other areas of the world are harder to come by, but it appears the trends hold elsewhere, with widespread multilingualism also reported in e.g. India and Indonesia. Overall, it seems reasonable to assume that more than half of the world's population speaks two or more languages.

To account for multilingualism in Gboard, users can enable on-the-fly, on-device mixing of many of Gboard's monolingual language models. In this multilingual mode, Gboard aims to automatically handle language switches by mixing these user-selected monolingual language models together to build a model that can account for multilingual usage. Because of keyboard layout constraints, it is not currently possible to enable this multilingual mode across languages that use different scripts.

### 5.4 Improving Quality Over Time

In general, the quality of a language model (and thus the quality of the input experience) depends on the size of the text corpus and on how well the corpus is matched to the target application domain. For example, a language model trained on news articles may show unexpected next-word predictions in conversational contexts, because it will rely on n-grams observed in news contexts.

More generally, for commonly written language varieties such as English (`en`), Russian (`ru`) and Chinese (`zh`), large text corpora can be found easily across many domains. This means that the typing experience upon first use will typically be better than in languages where smaller text corpora are available, with limited domain coverage. As described above, on-device personalization can help improve pre-built generic language models as the keyboard application is used over time, by creating a personal dictionary with out-of-vocabulary words and common phrases. In our user studies, we find that such on-device personalization usually helps improve the typing experience significantly.

Beyond on-device personalization, however, there are a number of ways in which these pre-built generic language models can be strengthened over time. Most obviously, as these language varieties become increasingly common on the public internet, web crawls can result in larger and more diverse text corpora.

Another approach is to apply a relatively new technique called "federated learning", where a distributed, privacy-preserving, on-device learning framework is used to train high-quality language models, as described in Hard et al. (2018) and Chen et al. (2019).

# 6 User Reactions

## 6.1 User Studies

After we've designed a layout and built a language model for a given language variety, we typically run a user study with a number of speakers of the target variety. This helps us make sure that the keyboard that's been built meets the needs of the community, which is critically important, as highlighted by Paterson (2015). These speakers are asked to install a beta version of the keyboard, and answer a survey with a variety of quantitative and qualitative questions to gauge their typing experience. This survey was originally conducted only in English, but we have since translated it into a number of other major world languages, such as French and Modern Standard Arabic, to make the survey easier to understand for communities around the world who may not be as familiar with English.

It would be impractical to present a full analysis of testers' responses across all language varieties here. In general, users express satisfaction at having full keyboard support for their language variety. Confirming our expectations at the start of this project, users frequently indicate they will use this keyboard to input their language in social media and chat apps in particular. Users typically also indicate they would be likely to recommend this keyboard app to others, with one of the top reasons being that it supports their own language variety.

Some testers also pointed out that having a keyboard application with support for their native language would help them feel more confident when typing, since they were still getting used to writing in their native language, and having features like auto-correction and spelling completions made it easier to spell their language correctly, since they weren't always entirely confident about the spelling of certain words.

Testers have, of course, also identified areas of improvement: most commonly, users indicate that the dictionary still appears to be relatively small, presumably arising from finding too many correctly spelled words being highlighted as spelling mistakes. This makes sense, given that the training corpora we trained the language models on are typically smaller than the corpora in other languages that our testers may be familiar with. Fortunately, on-device personalization can help address this by learning words over time as the keyboard is used.

Only in a handful of cases across the hundreds of user studies we have run so far did any testers indicate they would never write in their language variety. Even then, there was never agreement among the testers for a given language variety on this topic, with at least some testers indicating that they do use their language in writing, suggesting that perhaps writing is an emergent phenomenon in these language varieties.

To be fair, we did already exclude a handful of languages from our roadmap initially where we could find absolutely no evidence of any written tradition, despite them having a large number of speakers. Additionally, as described above, we prioritized our roadmap such that we work first on the languages where the impact is likely to be the highest. There is, therefore, certainly some amount of selection bias in these findings.

However, it has been exciting to see that, across many hundreds of languages, people indicate that a keyboard application with support for their language will make their everyday typing experience significantly easier, further affirming the theory that many of the world's languages are now increasingly used in writing to communicate in informal online environments, such as chat apps and social media.

These findings give us some cause for optimism: at least for the languages we have surveyed, which cover about 95% of the world's population by first language, the situation seems less bleak than was feared a few years ago, e.g. by Kornai (2013), who feared that "less than 5% of all languages can still ascend to the digital realm", and who suggested that perhaps at most 300 languages would see ever widespread usage online. Our surveys paint a somewhat more positive picture—although to be fair, we've only surveyed speakers of about 10% of the world's languages.

### 6.2 Usage Trends and Discoverability

We have generally seen an enthusiastic reception for most of the language varieties we have been releasing publicly. However, an interesting problem we have observed in follow-up interactions with speakers of these language varieties (who were not in the group of initial testers) is that many potential users remain unaware that their language is supported by their smartphone keyboard, even a year or so after it first became available.

This lack of awareness means speakers continue to use a sub-optimal input method for their variety, as described in the initial section showing how users solve for the limitations of language technology, even when those limitations have been removed. We have not done detailed studies in this area yet, but so far, it appears that there are a few factors at play. First, speakers may not have any expectation technology would support their language variety, because of a historic lack of support (of course, it would be unreasonable at any rate to expect most users to look at their phone's language settings menu on a regular basis). Second, even if users do look for their variety in the settings menu, these menus are complicated and challenging to less technically-savvy users.

In other cases, users are aware of their language variety being supported, but they express concerns that it will be hard to switch back to other language varieties that they commonly type in. Addressing these problems is an interesting area for future work: there may be technical solutions (e.g. creating an on-device system to automatically suggest the right language settings), but this is clearly non-trivial. In the meantime, it is worth noting here that again, we see problems that do not arise from the core underlying complexity of language technology, but more so from factors such as user interface design and user expectations of technology.

# 7 Conclusion

Many people without a linguistic background are surprised to learn that there are thousands of languages out there in the world. Even to those of us on the Gboard team with linguistic backgrounds, though, seeing the actual linguistic diversity of the world up close throughout our deep internationalization efforts for Gboard has been humbling. There is clearly an enormous linguistic diversity online these days, even if it is somewhat hidden from public view and awareness. In this context, localized keyboard applications can be of enormous benefit to users, as we have seen in many user studies across the world.

We hope that this technical report helps increase awareness of the needs and desires of language communities around the world, not just among developers of keyboard applications, but also more generally among practitioners in the field of language technology. We think there are bound to be more contexts in which product teams may be able to adopt the approaches we have developed to be effective in dealing with the unprecedented demands for scale in our keyboard software:

- designing natural-language processing algorithms and modeling approaches that scale to many languages, factoring in the world's linguistic diversity

- having clear, prioritized roadmaps for language support

- building automatic dashboards to track what has been done for each language, and what still needs to be done

- creating well-documented easy-to-follow processes for linguists and native speakers to inject knowledge

- using scalable infrastructure designed to work across any number of languages

- ensuring there are built-in mechanisms for feedback from language communities

It may seem like a daunting project to add support for hundreds of language varieties to a given product, but we hope we have shown it is possible to design scaleable approaches that help build products that support a very large number of languages. In our experience, it is especially helpful to do such design thinking before jumping in to doing any significant product development work, rather than designing a process that does work for a few dozen languages, but that does not scale to hundreds of languages.

Of course, many more language varieties remain that are not yet included in today's Gboard. And smartphone keyboards are just one application of language technology, with many other technologies, such as speech-to-text (voice dictation), still available in only dozens of languages. To achieve the goal of building for everyone, there is a lot of work ahead still for the field of language technology. But with the right approach in place, the focus can be on the interesting research, engineering and product challenges, as well as the rich human interactions, that are so critical to building technology for the world's languages.

## Acknowledgements

Many people have been involved in the Gboard deep internationalization effort in various roles over the years. First, we'd like to thank our fellow linguists in Google's Speech & Keyboard group, who have contributed to the research and engineering work for many of the language varieties supported in Gboard today: Alexa Cohen, Jonas Fromseier Mortensen, Shayna Lurya, Amanda Ritchart-Scott, Sandy Ritchie, Pierric Sans, and all the other linguists at Google who have helped us build these keyboards.

We'd also like to thank Caroline Kenny and Sarah Abu Sharkh for their invaluable efforts to support our data collection projects. Our thanks also go to Reena Lee, Angana Ghosh, and Linda Lin of the Gboard product management team, and all the members of the Gboard engineering team around the world.

Beyond the Gboard team, we'd also like to thank the members of Google's Internationalization Engineering team: Gboard could not function without their work on the Unicode libraries, the Google Noto fonts, and the Android text rendering stack. We would also like to thank the Speech/Language Algorithms team and our Language Model Infrastructure team for their help in extending our model training infrastructure so it could easily scale to hundreds of languages. Thanks, as well, to many other enthusiastic Googlers who have pitched in to help us with their own languages.

For providing inspiration and executive support, we thank Xu Liu, Pedro Moreno, and Johan Schalkwyk.

Finally, and most importantly, our deepest thanks go to all the linguists and native speakers outside of Google who have helped us create smartphone keyboards for their language varieties.